\documentclass[preprint,prl,twocolumn,10pt]{revtex4}%
\usepackage{amsfonts}
\usepackage{amsmath}
\usepackage{amssymb}
\usepackage{graphicx}%
\begin{document}
\title{Photon localization barrier can be overcome}
\author{P.~Saari$^{1,2}$, M.~Menert$^{1}$, and H.~Valtna$^{1}$}
\affiliation{$^{1)}$Department of Physics, University of Tartu; $^{2)}$Institute of
Physics, University of Tartu; Riia 142, Tartu 51014, Estonia}

\pacs{03.70.+k, 03.50.De, 03.65.Pm, 11.30.Cp}

\begin{abstract}
In contradistinction to a widespread belief that the spatial localization of
photons is restricted by a power-law falloff of the photon energy density,
I.Bialynicki-Birula [Phys. Rev. Lett. \textbf{80}, 5247 (1998)] has proved
that any stronger -- up to an almost exponential -- falloff is allowed. We are
showing that for certain specifically designed cylindrical one-photon states
the localization is even better in lateral directions. If the photon state is
built from the so-called focus wave mode, the falloff in the waist
cross-section plane turns out to be quadratically exponential (Gaussian) and
such strong localization persists in the course of propagation.

\end{abstract}
\maketitle

While quantum electrodynamics (QED) underwent an impressive development and
reached its maturity in the middle of the last century, one of its basic
concepts -- the photon wave function in free space -- was deprived of such
fortune. Although the photon wave function in coordinate representation was
introduced already in 1930 by Landau and Peierls~\cite{LP1930} the concept was
found to suffer from inherent difficulties that were not overcome during the
century (see review~\cite{IBBrev}). The common explanation presented in
textbooks (e.g.,~\cite{veneQED},\cite{MWrmt}) may be summed up as follows: (i)
no position operator exists for the photon, (ii) while the position wave
function may be localized near a space-time point, the measurable quantities
like the electromagnetic field vectors, energy, and the photodetection
probability remain spread out due to their non-local relation with the
position wave function. However, just before the turn of the century both of
these widely-espoused notions were disproved~\cite{HawtonPRA},\cite{IwoLoc}
and in the new century a fresh interest in the photon localization problem
seems to have been awakened (see, e.g.,~\cite{KellerPRA},\cite{KellerJOSAB}%
,\cite{EberlyPRL2002}), meeting the needs of developments in near-field
optics, cavity QED, and quantum computing.

I.~Bialynicki-Birula writes~\cite{IwoLoc} that the statement \textquotedblleft
even when the position wave function is strongly concentrated near the origin,
the energy wave function is spread out over space asymptotically like
$r^{-7/2}$ \textquotedblright\ (citation from~\cite{MWrmt}, p.~638) is
incorrect and that both wave functions may be strongly concentrated near the
origin. He demonstrates, on one hand, that photons can be essentially better
localized in space -- with an exponential falloff of the photon energy density
and the photodetection rates. On the other hand, he establishes -- and it is
even somewhat startling that nobody has done it earlier -- that certain
localization restrictions arise out of a mathematical property of the positive
frequency solutions which therefore are of a universal character and apply not
only to photon states but hold for all particles. More specifically, it has
been proven in the Letter~\cite{IwoLoc} for the case of spherically
imploding-exploding one-photon wavepacket that the Paley-Wiener theorem allows
even at instants of maximal localization only such asymptotic decrease of the
modulus of the wave function with the radial distance $r$ that is
\textit{slower} than the linear exponential one, i.e., anything slower than
$\sim\exp(-Ar)$, where $A$ is a constant. The latter is what the Paley-Wiener
theorem says about a function whose Fourier spectrum contains no negative frequencies.

The purpose of the present Letter is to indicate that one-photon wave
functions of a specific type can break the localization restriction and
exhibit the linear exponential and even faster falloff with the distance. Yet,
there is no contradiction either with the result of Ref.~\cite{IwoLoc} or with
the Paley-Wiener theorem, since the wave functions are cylindrical and exhibit
an exceptionally strong localization in two dimensions out of three.

As an introduction, we consider briefly the simplest case of a one-dimensional
Landau-Peierls wave function in order to indicate how the Paley-Wiener theorem
restricts the spatial localization of a photon. Then we study the radial
falloff for three different cylindrical wave functions, using exactly the same
formalism that has been presented in Ref.~\cite{IwoLoc}. Finally, the
discussion of our results allows us to refine the analysis given in
Ref.~\cite{IwoLoc}.

Let us consider a one-photon (1ph) state that corresponds to a plane-wave
pulse propagating unidirectionally, say, along the axis $z$, being polarized
along a lateral axis (say, the $x$ axis)%
\begin{equation}
\left\vert 1ph\right\rangle =\int_{0}^{\infty}dk_{z}~f(k_{z})a^{+}%
(k_{z})\left\vert vac\right\rangle ~, \label{1footvektor}%
\end{equation}
where $a^{+}(k_{z})$ is the creation operator of a photon and $f(k_{z})$ is a
properly-normalized photon wave function in the momentum representation. Then
the inverse Fourier transform (but including positive frequencies only!)
\begin{align}
\Phi(z,t)  &  =\frac{1}{2\pi}\int_{0}^{\infty}dk_{z}~f(k_{z})e^{i(k_{z}%
z-\omega t)}\label{1footLP}\\
&  =\frac{1}{2\pi}\int_{0}^{\infty}dk~f(k)e^{ik(z-ct)}%
\end{align}
represents the corresponding position space wave function of the photon in
state $\left\vert 1ph\right\rangle $ (see, e.g., \cite{MWrmt}, p.~636). The
modulus squared $\left\vert \Phi(z,t)\right\vert ^{2}$ gives the photon
probability density, i.e., the degree of localization along the axis $z$ (in
the given case in the directions $x$ and $y$ any localization is absent). If
$f(k)$ differs from zero within a wide frequency band, the probability
$\left\vert \Phi(z,t)\right\vert ^{2}$ may be strongly localized around a
point $z_{0}$ moving along the axis $z$ with the speed of light $c.$ However,
since $\Phi(z,t)$, due to the absence of negative frequencies in the integral
of Eq.(\ref{1footLP}), is nothing but a complex analytic signal, according to
the Paley-Wiener theorem (or criterion) the asymptotic decrease of $\left\vert
\Phi(z,t)\right\vert ^{2}$ with the distance $r=|z-z_{0}|$ has to be weaker
than $\sim\exp(-Ar)$, where $A$ is a constant. All the more excluded are any
finite-support functions in the role of $\Phi(z,t)$. To conclude the
introduction, let us notice that if the counterpropagating (with $k_{z}<0$)
Fourier components are involved in Eq.(\ref{1footLP}), the Paley-Wiener
theorem does not apply at the instant $t=0$.

Following Ref.~\cite{IwoLoc}, we shall study the photon localization by
examining the asymptotic behavior of the positive frequency part of the
Riemann-Silberstein vector $\mathbf{F}(\mathbf{r},t)$ (called the energy wave
function) which directly determines the energy density of a one-photon state
and is conveniently expressed through a "superpotential" $\mathbf{Z}%
(\mathbf{r},\tau)$%
\begin{equation}
\mathbf{F}(\mathbf{r},t)=\nabla\times\left[  i\frac{\partial}{\partial\tau
}\mathbf{Z}(\mathbf{r},\tau)+\nabla\times\mathbf{Z}(\mathbf{r},\tau)\right]
~, \label{FviaZ}%
\end{equation}
where $\tau\equiv ct$. The vector field $\mathbf{Z}(\mathbf{r},\tau)$ is
nothing but an analytic signal version of the Hertz potential, i.e.,
$\mathbf{Z}(\mathbf{r},\tau)=\mathbf{m}\Psi(\mathbf{r},\tau),$ where
$\mathbf{m}$ is a constant vector that includes the proper normalization
factor and $\Psi(\mathbf{r},\tau)$ is any solution of the scalar wave
equation, which is taken in the form of the analytic signal.

As the first example leading to a stronger localization that one might expect
from the Paley-Wiener theorem, let us consider the photon field where
$\mathbf{m}$ is directed along the axis $z$ (any other orientation gives
similar results) and $\Psi(\mathbf{r},\tau)$ is a superposition of
cylindrically symmetric Bessel functions $J_{0}$ as a wavepacket with the
exponential spectrum and a specific dispersion law for the axial wavenumber
$k_{z}(\omega)=const=k_{0}$%
\begin{equation}
\Psi(\mathbf{\rho},z,\tau)=\int_{|k_{0}|}^{\infty}dk\ J_{0}\left(  k_{\rho
}\rho\right)  e^{-k\Delta}e^{-i(k\tau-k_{0}z)}~,\label{SilindriIntegr}%
\end{equation}
where the radial coordinate $\rho$ has been introduced and $k_{\rho}=\left(
k^{2}-k_{0}^{2}\right)  ^{1/2}$ is the lateral component of the wave vector of
the monochromatic plane-wave constituents represented with the weight function
$e^{-k\Delta}$ whose width is $\Delta^{-1}.$ The integral can be taken with
the help of a Laplace transform table and we obtain%
\begin{equation}
\mathbf{Z}(\rho,z,\tau)=\mathbf{m}\frac{\exp\left(  -\left\vert k_{0}%
\right\vert \sqrt{\rho^{2}+\left(  \Delta+i\tau\right)  ^{2}}\right)  }%
{\sqrt{\rho^{2}+\left(  \Delta+i\tau\right)  ^{2}}}e^{ik_{0}z}%
~.\label{YldSilinderLaine}%
\end{equation}
Eq.(\ref{YldSilinderLaine}) describes a simple cylindrical pulse modulated
harmonically in the axial direction and radially converging (when $\tau<0$) to
the axis and thereafter (when $\tau>0)$ expanding from it, the intensity
distribution resembling an infinitely long tube coaxial with the $z$ axis and
with a time-dependent diameter (see Fig.~6. in Ref.~\cite{LorTransPRE}). It
follows from Eqs.~(\ref{YldSilinderLaine}) and (\ref{FviaZ}) that%
\begin{align}
|\mathbf{Z}(\rho &  \rightarrow\infty,z,\tau=0)|\sim\rho^{-1}\exp
(-\rho/l)~,\label{SilindriFalloffA}\\
|\mathbf{F}(\rho &  \rightarrow\infty,z,\tau=0)|^{2}\sim\left[  \rho
^{-2}+O(\rho^{-3})\right]  \exp(-2\rho/l)~,\label{SilindriFalloffB}%
\end{align}
where $l\equiv$ $\left\vert k_{0}\right\vert ^{-1}$ is the characteristic
length (or length unit). Thus, while the photon is delocalized in the axial
direction, its energy density falloff in the lateral directions is exactly the
linear exponential one at all times the conditions $\tau\ll\rho\gg\Delta$ are
fulfilled, see Fig.~1. The time derivative as well as the spatial derivatives
contain the same exponential factor, ensuring the exponential falloff of the
the Riemann-Silberstein vector in Eq.(\ref{SilindriFalloffB}).
\begin{figure}
[ptbh]
\begin{center}
\includegraphics[
height=7.6573cm,
width=6.2893cm
]%
{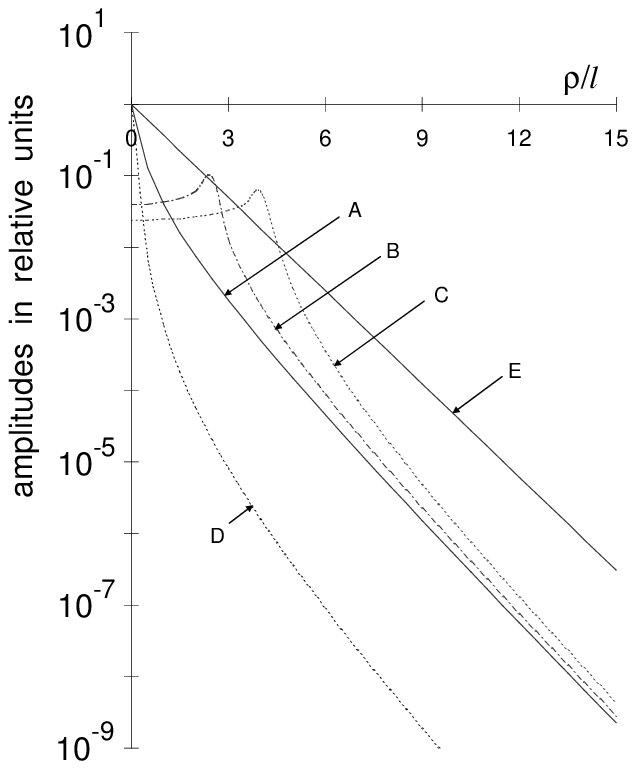}%
\caption{Curves of the radial dependence in a decimal logarithmic scale. Curve
A is for $\left\vert \mathbf{Z}(\rho,0,\tau=0)\right\vert $; B, $\left\vert
\mathbf{Z}(\rho,0,\tau=2.5l)\right\vert $; C is the same as B but with $\Psi$
taken from Eq.~(\ref{FXWminu}); D, $\left\vert \frac{\partial}{\partial\tau
}\mathbf{Z}(\rho,0,\tau=0)\right\vert $; E is a reference curve $\exp
(-\rho/l)$. The curves A, B, and C have been normalized so that $\left\vert
\mathbf{Z}(0,0,0)\right\vert =1.$ The values of the remaining free parameters
are $\Delta=0.1l$ and $\beta=0.8$.}%
\end{center}
\end{figure}
Hence, a one-photon field given by Eq.(\ref{YldSilinderLaine}) serves as the
first and simplest example where the localization in two transversal
dimensions is governed by different rules than localization in three
dimensions according to Ref.~\cite{IwoLoc}.

The next example is readily available via the Lorentz transformation of the
wave function given by Eq.(\ref{SilindriIntegr}) along the axis $z$, which
gives another possible solution of the scalar wave equation. The result is a
new independent solution but it can also be considered as the wave given by
Eqs.~(\ref{SilindriIntegr}) and (\ref{YldSilinderLaine}), which is observed in
another inertial reference frame~\cite{LorTransPRE}:%
\begin{align}
\Psi(\mathbf{\rho},z,\tau) &  =\frac{\exp\left(  -\left\vert k_{0}\right\vert
\sqrt{\rho^{2}+\left(  \Delta-i\gamma\left(  \beta z-\tau\right)  \right)
^{2}}\right)  }{\sqrt{\rho^{2}+\left(  \Delta-i\gamma\left(  \beta
z-\tau\right)  \right)  ^{2}}}\nonumber\\
&  \qquad\times\exp\left(  i\gamma k_{0}\left(  z-\beta\tau\right)  \right)
\ ,\label{FXWminu}%
\end{align}
where the relativistic factors $\gamma\equiv(1-\beta^{2})^{-1/2}$ and
$\beta\equiv v/c<1$ have been introduced, $v$ being a free parameter -- the
relative speed between the frames. In the waist region (see Fig.~2) this wave
function has the same radial falloff as was given by
Eq.(\ref{SilindriFalloffA}), see curve "C" in Fig.~1, while the axial
localization follows a power law. The strongly localized waist and the whole
amplitude distribution move rigidly and without any spread along the axis $z$
with a superluminal speed $c/\beta.$ Such wave with intriguing properties,
named the focused X wave (FXW)~\cite{revPIER}, belongs to the so-called
propagation-invariant localized solutions to the wave equation -- a research
field emerged in the 1980-ies (see reviews \cite{revPIER}-\cite{KaidoPhD}) and
recently reached its first experimental results \cite{PRLmeie}-\cite{exp4Lisa}%
.%
\begin{figure}
[ptbh]
\begin{center}
\includegraphics[
height=6.8513cm,
width=8.7444cm
]%
{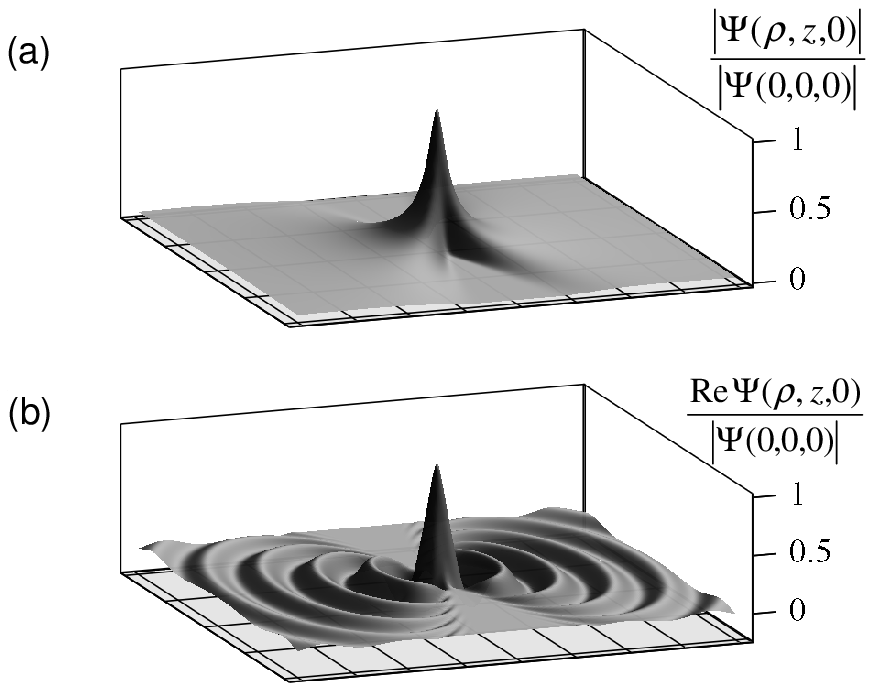}%
\caption{The superluminal FXW given by Eq.~(\ref{FXWminu}). Shown are the
dependences (a) of the modulus and (b) of the real part of the wavefunction on
the longitudinal ($z,$ increasing to the right) and a transverse (say, $x)$
coordinates. The distance between the grid lines on the basal plane $(x,z)$ is
$22\lambda$, where $\lambda=2\pi|k_{0}|^{-1}$, $k_{0}$ being negative. The
values of the remaining free parameters are $\Delta=$ $30\lambda$ and
$\beta=0.995$ or $\gamma=10$.}%
\end{center}
\end{figure}
Hence, in its waist (cross-sectional) plane a one-photon field given by the
FXW possesses the same strong localization at any time as the previously
considered cylindrical field does in any transversal plane at the instant
$t=0$.

By making use of the historically first representative of localized waves --
the so-called focus wave mode (FWM)~\cite{FWMpioneer1}-\cite{LWpioneere4} (see
also Ref.~\cite{LorTransPRE} and reviews~\cite{revPIER}, \cite{KaidoPhD} and
references therein) one readily obtains an example of the field that exhibits
even much stronger than exponential localization. FWM is given by the scalar
function%
\begin{equation}
\Psi(\rho,z,t)=\frac{\exp\left[  -\frac{\rho^{2}}{2l\left(  a-i\left(
z-\tau\right)  \right)  }\right]  }{a-i\left(  z-\tau\right)  }\exp\left[
-\frac{i\left(  z+\tau\right)  }{2l}\right]  , \label{FWmode}%
\end{equation}
where again $l$ is a wavelength-type characteristic length and the constant
$a$ controls the axial localization length. Since the FXW in the limit
$\beta\rightarrow1$ becomes a FWM \cite{LorTransPRE}, Fig.2 gives also an idea
how a FWM looks like. Multiplying Eq.(\ref{FWmode}) by $\mathbf{m}$ to build
the vector $\mathbf{Z}(\rho,z,\tau)$ and inserting the latter into
Eq.(\ref{FviaZ}) we obtain that in this example the photon localization in the
waist plane is quadratically exponential (Gaussian falloff):%
\begin{align}
|\mathbf{Z}(\rho &  \rightarrow\infty,z=\tau)|\sim\exp(-\rho^{2}%
/2la),\label{FWMfalloffA}\\
|\mathbf{F}(\rho &  \rightarrow\infty,z=\tau)|^{2}\sim\rho^{6}\exp(-\rho
^{2}/la)~. \label{FWMfalloffB}%
\end{align}
In Eq.(\ref{FWMfalloffB}) only the highest-power term with respect to $\rho$
is shown.

To start discussing our results let us ask first whether the wave functions
considered are something extraordinary. The answer is: yes, they are indeed,
since the browsing of various integral transform tables reveals rather few
examples where both the real and imaginary part of a wave function and of its
time derivative have simultaneously an exponential or stronger localization in
conjunction with other requisite properties. Still, the list of proper wave
functions with an extraordinary strong localization is not poor -- in addition
to an optically feasible version~\cite{meieBesselGauss} of the FWM various new
interesting solutions can be derived~\cite{KiseleviYld}. Yet, it could be
argued that the well-known Gaussian beam pulse has the same quadratically
exponential radial profile in the waist region. However, resorting to the
family of the Gaussian beams (the Gauss-Laguerre and Gauss-Hermite beams,
etc.) is irrelevant here. The reason is that all these beams are solutions of
the wave equation only in the paraxial approximation not valid in the case of
any significant localization of wide-band (pulsed) superpositions of the
beams, whereas in fact, e.g., an exact solution corresponding to a
lowest-order (axisymmetric) Gaussian beam has a weak power-law radial falloff
in the waist region~\cite{ColiniEMBeyondParax},\cite{MPSminu}.

The next possible objection to the physical significance of the results
obtained might arise from the infinite total energy \cite{revPIER} of the
waves given by Eqs.~(\ref{YldSilinderLaine}), (\ref{FXWminu}), and
(\ref{FWmode}). However, at any spatial location the wave function is square
integrable with respect to time, thus the condition of the Paley-Wiener
theorem has been satisfied. Moreover, physically feasible finite-energy
versions of localized waves generally exhibit even better localization
properties, although not persistently. A finite-energy version of the FXW,
called the modified focused X wave (MFXW~\cite{revPIER}), has the same
exponential factor as in Eq.(\ref{FXWminu}), which is multiplied by a fraction
that allows to force the axial localization to follow an arbitrarily strong
power-law. The latter circumstance indicates that the strong lateral
localization of the fields considered does not appear somehow at the expense
of their axial localization. As a matter of fact, energy-normalization of a
wave function depends on how many photons it describes. It is easy to see that
our derivation and results hold for any number state with $N\geq1$ and also
for incoherent mixtures of such states (which is important for experimental
studies). In contrast, the localization of coherent states is not governed by
the Paley-Wiener theorem~\cite{IwoLoc}.

The final crucial question is, are our results not in contradiction with those
of Ref.~\cite{IwoLoc}? The answer is no, since in the case of the cylindrical
waves the radial distance and temporal frequency are not directly
Fourier-conjugated variables. In order to clarify this point, let us first
take a closer look at the proof of the Paley-Wiener limit for
three-dimensional isotropic localization. In Ref.~\cite{IwoLoc} Eq.(24) for
$\mathbf{Z}(\mathbf{r},\tau)$ contains a superposition of spherically
symmetric standing waves%
\begin{equation}
\int_{0}^{\infty}dk~h(lk)\frac{\sin kr}{r}e^{-ik\tau}\propto ir^{-1}\left[
g(\frac{\tau+r}{l})-g(\frac{\tau-r}{l})\right]  ~, \label{IBBeq24}%
\end{equation}
where $h(lk)$ is the spectrum and $g(.)$ is its Fourier image. The sine in
Eq.(\ref{IBBeq24}) results from the imploding and exploding spherical wave
constituents of the standing wave, like an odd one-dimensional standing wave
arises from counterpropagating waves. While the asymptotic behavior of the
function $g(.)$ and hence of the function $\mathbf{Z}(\mathbf{r},\tau)$ for
large values of the radial distance $r$ are generally restricted by the
Paley-Wiener theorem, strictly at the instant $\tau=0$ of maximal localization
the integral is nothing but the sine transform for which the theorem does not
apply. Indeed, the sine transform tables give examples of the resultant
functions with arbitrarily abrupt falloff. However, it does not mean as if the
localization restriction was lifted at the instant $\tau=0$. The explanation
is that according to Eq.(\ref{FviaZ}) the energy wave function involves also
the time derivative of $\mathbf{Z}(\mathbf{r},\tau)$, but the sine transforms
of two functions $h(lk)$ and $h(lk)k$ cannot simultaneously possess
arbitrarily abrupt falloffs. In contrast, the time derivative of the wave
function given by Eq.(\ref{YldSilinderLaine}) or Eq.(\ref{SilindriIntegr}) has
the same strong exponential falloff as the function itself, which persists for
some (not too long) time, see Fig.1. By comparing Eqs.~(\ref{SilindriIntegr})
and (\ref{IBBeq24}) we notice that while in Eq.(\ref{IBBeq24}) -- as well as
in its one-dimensional equivalent -- the argument of the sine function is the
product of the distance with the Fourier variable, in Eq.(\ref{SilindriIntegr}%
) the argument of the Bessel function is the product of the radial distance
$\rho$ with the radial wavenumber $k_{\rho}$ the latter depending on the
Fourier variable through the square-root expression with the constant
parameter $k_{0}$ -- the lower limit of the integration. As it follows also
from Eqs.~(\ref{SilindriFalloffA}) and (\ref{SilindriFalloffB}) the condition
$k_{0}\neq0$ is crucial for obtaining the exponential falloff. Hence, in the
case of the cylindrical waves considered by us, the apparent violation of the
rules set by the Paley-Wiener theorem results from the specific complicated
relation between the radial distance and the Fourier variable.

In conclusion, we have shown that for certain cylindrical $N$-photon states
($N=1,2,...$) the localization in lateral directions breaks the limit
established in Ref.~\cite{IwoLoc} for the case of uniform spherical wave
functions. These results hold not only for photons but for number states of
any particles.

The first author is thankful to Iwo Bialynicki-Birula for stimulating hints,
discussions, and remarks. The research was supported by the Estonian Science Foundation.

\end{document}